# Wide range temperature-dependent (80–630 K) study of Hall effect and the Seebeck coefficient of $\beta$-Ga$_2$O$_3$ single crystals

Ashish Kumar[1,*], Saurabh Singh[2], Bhera Ram Tak[3], Ashutosh Patel[4], K. Asokan[1], D. Kanjilal[1].

[1]*Inter-University Accelerator Centre, Vasant Kunj, New Delhi-110067, India*
[2]*Toyota Technological Institute, Hisakata 2-12-1, Tempaku, Nagoya 468-8511, Japan.*
[3]*Department of Physics, Indian Institute of Technology Delhi, Hauz Khas, New Delhi-110019, India.*
[4]*Department of Mechanical Engineering, IISc Bangalore-560012, India.*

*email address: ashish@iuac.res.in, dr.akmr@gmail.com*

**Abstract:**
Investigation of Seebeck coefficient in ultra-wide bandgap materials presents a challenge in measurement, nevertheless, it is essential for understanding fundamental transport mechanisms involved in electrical and thermal conduction. $\beta$-Ga$_2$O$_3$ is a strategic material for high power optoelectronic applications. Present work reports Seebeck coefficient measurement for single crystal Sn doped $\beta$-Ga$_2$O$_3$ in a wide temperature range (80- 630 K). The non-monotonic trend with large magnitude and negative sign in the entire temperature range shows electrons are dominant carriers. The structural and Raman characterization confirms the single-phase and presence of low, mid, and high-frequency phonon modes, respectively. Temperature dependent (90-350 K) Hall effect measurement was carried out as supplementary study. Hall mobility showed $\mu \propto T^{1.12}$ for T<135 K and $\mu \propto T^{-0.70}$ for T>220 K. Activation energies from Seebeck coefficient and conductivity analysis revealed presence of inter band conduction due to impurity defects. The room temperature Seebeck coefficient, power factor and thermal conductivity were found as 68.57 ±1.27 μV/K, 0.15 ±0.04 μW/K$^2$cm and 14.2 ± 0.6 W/mK, respectively. The value of the *figure-of-merit* for $\beta$-Ga$_2$O$_3$ was found to be ∼ 0.01 (300 K).

Ga$_2$O$_3$ is being explored for high power and voltage devices, transparent electronics (phosphors and electroluminescent display), catalysis, gas sensors, etc. largely due to its excellent semiconducting properties like wide bandgap (4.7 – 4.9 eV), and small electron effective mass (0.25 – 0.28 m$_e$) [1-4]. $\beta$-monoclinic phase of Ga$_2$O$_3$ polymorph is highly stable in temperature and radiation harsh environments [5, 6]. Its electrical and thermal conductivity properties are extensively studied, however, thermoelectricity is rarely explored, mainly because of difficulty in the measurement of the Seebeck coefficient for ultra-wide band gap materials [7-10]. Information about Seebeck coefficient for a material is essential for exploring potential applications by the evaluation of *power factor* ($P = S^2\sigma$), and *figure of merit* ($ZT = S^2\sigma T/\kappa$), where $S$, $\sigma$, $\kappa$, and $T$ are Seebeck coefficients, electrical conductivity, thermal conductivity, and absolute temperature, respectively. Also, study of temperature dependence of Seebeck coefficient and Hall effect helps to understand various dominant scattering mechanisms, the density of states, effective mass, shallow and deep energy levels, the shift of chemical potential, and types of defects present in the semiconductor material [11, 12]. For wide bandgap semiconductors (GaN, Ga$_2$O$_3$, etc.) accurate measurement of the Seebeck coefficient is relatively a tricky exercise due to obstacles such as non-ohmic contacts, very low conductivity ($\sigma$), etc. Further, from the theoretical point of view, thermopower studies in wide bandgap semiconductors are affected by underestimated bandgaps, mainly due to the limitations of the approximations used in the calculations of the electronic structure. Further, the high thermal conductivity (10-30 W/mK) of $\beta$-Ga$_2$O$_3$ is one of the major issues which restrict this material for commercial thermoelectric applications in the present form. Thermal conductivity of similar monoclinic crystal structure in chalcogenides materials such as Ag$_2$(S, Se, Te) possesses very low thermal conductivity (less than 1 W/m$^{-1}$K$^{-1}$) which is attributed to the an-harmonic scattering and disorder structure [13, 14]. In these materials, it is expected that the thermal conductivity might be decreased by introducing the disorder with the short and long-wavelength phonon scatterings for both low and high-temperature regimes. Similarly, in GaN controlled defect introduction by swift heavy ion irradiation have resulted in increased thermopower and power factor [15]. An optimized ion energy and fluence can improve the thermoelectric properties of $\beta$-Ga$_2$O$_3$. Interestingly, the recent reports show that introducing a small amount of Ga$_2$O$_3$ as a secondary phase into the In$_2$O$_3$(ZnO)$_3$ and ZnO materials, the thermal conductivity drastically decreases due to enhancement in the scattering of long-wavelength phonons [16, 17]. The mechanical hardness and electrical conductivities are also found to be significantly changed by the addition of $Ga_2O_3$ in a state of art thermoelectric materials [18, 19]. For the bulk device, composites and polycrystalline materials are suitable. However, for the energy harvesting or cooling applications in on-chip devices, thin-film or single crystals have more advantages due to the size and compatibility. In a very recent study, Boy et al. reported the Seebeck coefficient of homo-epitaxial thin films of Si-doped *β*-$Ga_2O_3$ relative to Al [9]. However, no such reports on bulk or single-crystal samples exists to the best of our knowledge. Therefore, it is imperative to study *β*-$Ga_2O_3$ thermoelectric for a better understanding of transport properties as well as their potential applications.

In this letter, we present the results of temperature-dependent Hall-effect and Seebeck measurements on bulk *β*-$Ga_2O_3$ grown by melt growth method. A detailed analysis has been carried out to understand the thermoelectric behavior in a wide temperature range (80–630 K).

The single crystals of Sn doped *β*-$Ga_2O_3$ (thickness 680 μm) grown by the melt growth EFG (Edge-defined Film-fed Growth) method were procured from M/S Tamura Corp., Japan. The structural analysis of this single crystal was performed using Philips X'pert Pro X-ray diffraction (XRD) system with Cu-$K_\alpha$ radiation of wavelength 1.54 Å. The Raman measurements (LabRAM HR Horiba Jobin Yvon) were carried out with 532 nm excitation wavelength. Circular Ohmic contact pads of Ti/Au (20/80 nm) bilayer were fabricated in the cleanroom environment using an ultra-high vacuum electron beam evaporation system [20]. The circular contact pads of 0.5 mm diameter were fabricated at the two opposite edges of rectangular samples (2×5 mm$^2$) for thermoelectric measurements, and at four corners of square samples (5×5 mm$^2$) in van der Pauw geometry for electrical characterizations (separate pieces from the same wafer). The Seebeck coefficient measurement was performed in the temperature range of 80–630 K using an in-house developed system [21]. A pre-measurement calibration of the system using a constantan standard was done. Similarly, temperature-voltage and current-voltage measurements were performed at room temperature to ensure good contact between probes and samples in the Seebeck and Hall Effect measurements (90–350K), respectively. Room temperature thermal conductivity measurement was carried out for the *figure-of-merit calculation* of $Ga_2O_3$ material.

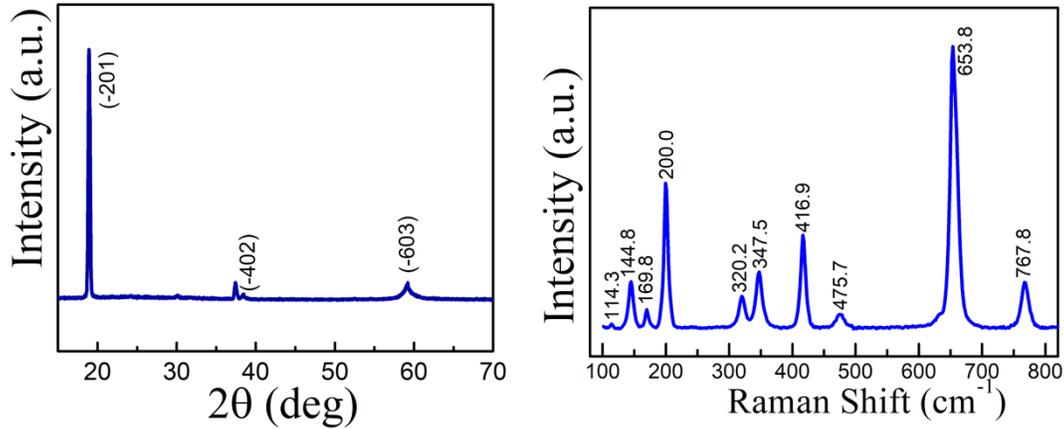

*Figure 1 X-ray diffraction (a) and Raman spectrum (b) of ($\bar{2}01$) oriented monoclinic β-Ga$_2$O$_3$ single crystal.*

The XRD (2-theta) scan of β-Ga$_2$O$_3$ single crystal is shown in figure 1a. The highest intense peak was observed at 18.93° which is corresponding to the ($\bar{2}01$) plane (JCPDS No. - 431012). All higher-order diffraction peaks of planes ($\bar{4}02$) and ($\bar{6}03$) were also observed. The XRD analysis revealed the monoclinic structure and ($\bar{2}01$) orientation with lattice parameters $a$=12.23 Å, $b$=3.04 Å and $c$=5.8 Å which were found to be in very good agreement with data provided by the material supplier.

The micro Raman spectrum of single-crystal Ga$_2$O$_3$ was recorded and is shown in figure 1b. In β-Ga$_2$O$_3$, only 15 phonon modes are Raman active among 27 modes at γ-point [22]. Here, 10 Raman modes have been observed from the spectral range of 100 to 900 cm$^{-1}$ that are in good agreement with the reported phonon modes of monoclinic β-Ga$_2$O$_3$ [22, 23]. These phonon modes are divided into three broad-spectral regions. The low-frequency modes below 200 cm$^{-1}$ originate due to translation motions of tetrahedron and octahedron chains. The origin of mid-frequency phonon modes between 300-500 cm$^{-1}$ are from the deformation of Ga$_2$O$_6$ octahedra whereas the high-frequency modes above 500 cm$^{-1}$ are related to stretching and bending of GaO$_4$ tetrahedra. In our case, four low-frequency, four mid-frequency, and two high-frequency Raman modes are evident for β-Ga$_2$O$_3$ single crystal.

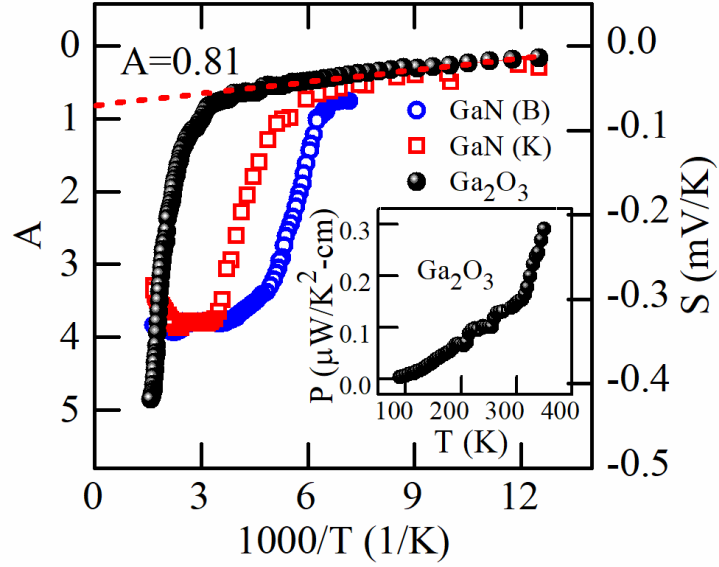

*Figure 2 Experimental Seebeck coefficients of the bulk of β-Ga2O3 measured in the present study plotted with n-GaN thin films for comparison. The reported data by Brandt et al. (legend GaN (B))[24] and Kumar et al. (legend GaN (K))[21] show a similar trend. Extrapolation of S vs. 1000/T for data in (80–300K) region on the Y-axis (left) gives a scattering factor (A) of 0.81. The inset shows the power factor for β-Ga$_2$O$_3$ measured as a function of temperature.*

Figure 2 shows the temperature dependence (1000/T) of the Seebeck coefficient (right Y-axis) of Sn doped β-Ga$_2$O$_3$. Two distinct regions with different slopes above and below of 300 K can be seen. Similar temperature dependence behavior has also been reported in other wide bandgap semiconductors like GaN [21, 24] having similar carrier concentrations ($n=2\times10^{18}$ cm$^{-3}$). The power factor (P) vs temperature curve is shown in the inset of figure 2. The thermal conductivity (κ) at room temperature was also measured for completeness of investigation and was found to be 14.2 ± 0.6 W/mK and is close to the reported range for bulk β-Ga$_2$O$_3$ [3]. The value of the *figure-of-merit ZT* is found to be ~ 0.01 (300 K), which is very low as compared to the commercially used thermoelectric materials.

Fundamentally, the Seebeck coefficient is a measure of mean energy carried w.r.t. Fermi level ($E_F$) by the charge carriers when a temperature gradient is applied. Under the energy (E) relaxation time (τ) approximation $\tau \propto E^r$, the Seebeck coefficient for negative charge carriers is given by Boltzmann equation as [11, 12]

$$S = -\frac{k}{e}\left[A + \frac{\Delta E_{TP}}{kT}\right] \quad (1)$$

where $\Delta E_{TP}$ is the activation energy for thermopower. $A$ (= 5/2 + $r$) is the scattering factor and is assumed constant for measured temperature range while the rest of the symbols have usual meanings. In comparison, the activation energy ($\Delta E_\sigma$) for conductivity can be determined as [11, 12]

$$\sigma = \sigma_o \exp\left[\frac{\Delta E_\sigma}{kT}\right] \qquad (2)$$

where $\sigma_o$ is constant. The difference of two activation energies, $\Delta E = \Delta E_\sigma - \Delta E_{TP}$ can be used to estimate the size of potential fluctuations at the conduction band edge created by the charged impurities and defects as explained in cited references [24-26]. The activation energy of thermopower ($\Delta E_{TP}$) and the scattering factor ($A$) calculated from slope and intercept (80 –300 K) were 4.66 ±0.14 meV and 0.81 ± 0.01, respectively (see table 1). Since the measured room temperature concentration $n=2.15\times10^{18}$ cm$^{-3}$ was less than the density of states in conduction band $N_C=3.72\times10^{18}$ cm$^{-3}$, the use of the non-degenerate Boltzmann equation (eq.1) under energy relaxation time approximation ideally should give scattering factor ($A$) values between 2 (scattering acoustic phonons) and 4 (scattering by ionized impurities). However, the observed value of $A$ (0.81) around room temperature was lower than the scattering by acoustical phonons ($A=2$), indicating different charge transport and scattering mechanisms were involved.

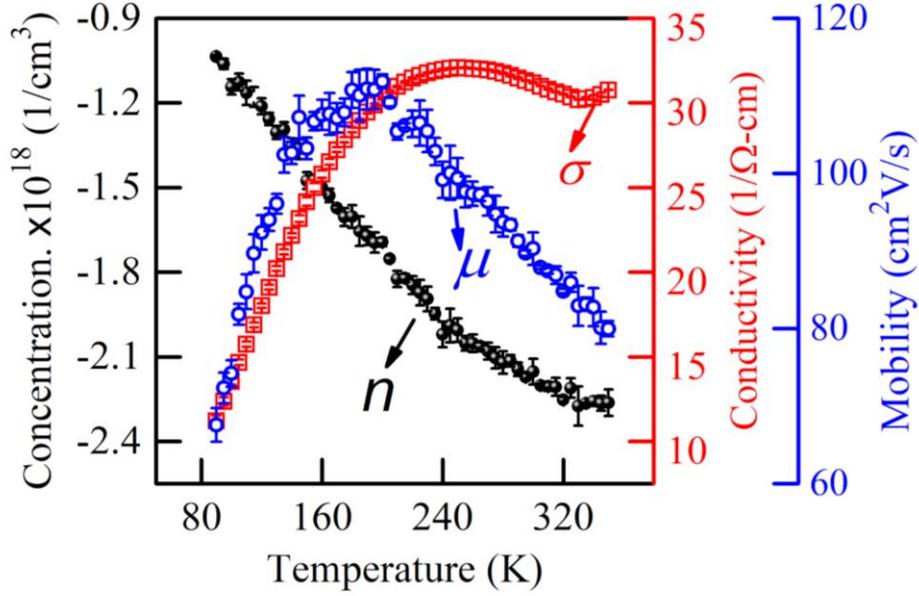

*Figure 3: Hall effect measurement data in the temperature range (90 – 350K). Calculated activation energies for conductivity (σ) and carrier concentration (n) are listed in table 1. Mobility (µ) is maximum at 190 K and shows $T^{1.12}$ and $T^{-0.70}$ dependence for region T<135 K and T>220 K, respectively.*

To account for the scattering factor, the Hall mobility measurement was performed in 90–350 K temperature range (figure 3), which shows typical mobility curve with *µ* values 90 cm$^2$V/s at 300 K and a maxima 110.86 cm$^2$V/s at 190 K. Interestingly, the mobility curve shows $\mu \propto T^{1.12}$ for the low-temperature region (T<135 K) and $\mu \propto T^{-0.70}$ for the high-temperature region (T>220 K). The $\mu \propto T^{-0.70}$ lies between $T^{-0.50}$ (polar optical phonon scattering) and $T^{-1.5}$ (for acoustic phonon scattering), indicating the presence of strong polar optical scattering around room temperature. Na *et al*. had attributed this mobility dependence on dominant Frohlich interaction caused by high ionicity of Ga-O bonds along with the low optical phonon energies in non-degenerate concentrations [8]. Oishi *et al.* found a high level of impurities in EFG grown samples which formed the interband in the energy bandgap and resulting similar mobility dependence on temperature [27]. For temperatures <150 K, the scattering from ionized and neutral impurity dominates. The carrier concentration shows a continuous increase for the measured temperature range as more electrons from the dopant level migrate to the conduction band. The conductivity variation observed in this study is similar to the report of Oishi et al. based on the Hall effect measurement of *β*-Ga$_2$O$_3$ [27]. The activation energy from the Hall concentration plot gives $E_n = $ 10.17 ±0.17 meV. The standard semiconductor statistics can be used to determine the dopant (Sn) donor level ($E_d$) in $\beta$-Ga$_2$O$_3$ from the activation energy of Hall concentration ($E_n$), which gives $E_d$ = 2 $E_n$ for no compensation and $E_d = E_n$ when compensation of carriers takes place. The donor level of Sn from this study is consistent with the reported [28-31] values from 7.4 meV to 60 meV for Sn doping.

*Table 1 Calculated experimental parameters from Figure 1 and 2. Activation energies and room temperature constants are determined by fitting the experimental data plotted as the Arrhenius equation.*

| Sample Parameters | $\beta$-Ga$_2$O$_3$ (Experimental) |
|---|---|
| $S$ (μV/K) | 68.57 ±1.27 (300K) |
| $P$ (μW/K$^2$cm) | 0.15 ±0.04 (300 K) |
| $\sigma$ (1/Ω cm) | 31.18 ±0.03 (300K) |
| $n$ (cm$^{-3}$) | 2.15E-18 ±0.04E-18 (300K) |
| $\mu$ (cm$^2$V/s) | 90.30 ±2.10 (300K) |
|  | 110.86 ±2.63 (at 190 K) |
| $\Delta V$ (meV) | 9.79 (300K) |
| $A$ | 0.81 ± 0.01 (80 -300K), |
|  | 12.92 ±0.31(400-600K) |
| $\Delta E_{TP}$ (meV) | 4.66 ±0.14 (80-300K), |
|  | 456.09 ±13.9 (400-600K) |
| $\Delta E_\sigma$ (meV) | 14.87 ±0.05 (80-220K) |
| $\Delta E = \Delta E_\sigma - \Delta E_{TP}$ (meV) | 10.21 (80-300K) |
| $E_n$ (meV) | 10.17 ±0.17 (80-220K) |

To understand the charge transport mechanism in wide bandgap materials, analysis of activation energies from the Hall effect and the thermopower experiments have been used [24, 25]. For an ideal band to band conduction, activation energy ($\Delta E_\sigma$) due to the conductivity must be larger than the activation energy of thermopower ($\Delta E_{TP}$). Observed low to room temperature (80−300 K) $\Delta E_\sigma$ (14.87±0.05 meV) in the present study, is greater than $\Delta E_{TP}$ (4.66±0.14). The difference of two activation energies ($\Delta E = \Delta E_\sigma - \Delta E_{TP}$ = 10.21 meV) is a measure of potential fluctuations at band edges. Such fluctuations can also be estimated by the approach given by Kane *et al.* [26], which can confirm our activation energy analysis.

$$\Delta V = \frac{e}{4\pi\varepsilon\varepsilon_o} \left[ \left( \frac{\varepsilon\varepsilon_o kT}{ne^2} \right)^{1/2} N_d \right]^{1/2} \quad (3)$$

Here, the symbols have their usual meanings. Assuming complete activation of donor concentration $N_D$ into charge carriers i.e. $N_D = n = 2.15 \times 10^{18}$ cm$^{-3}$ at room temperature (T= 300 K), and relative permittivity at $\varepsilon = 10$ [3], the value of $\Delta V$ is found to be 9.79 meV and is in close agreement with $\Delta E$ determined earlier, suggesting fluctuations are present but the presence of secondary hopping transport paths at low temperature could not be confirmed which has been commonly reported in disordered semiconductors and other wide bandgap semiconductors [24, 25]. However, for high temperatures (400−600K), $\Delta E_{TP}$ increased substantially (456.09 ±13.9 meV), which can be an indicator of presence of hopping or other secondary conduction from defect states in the bandgap of $\beta$-Ga$_2$O$_3$. The observed conduction process in the Hall effect and $S$ measurements collectively can be divided into two temperature regimes:

(i) Low to room temperature: here, Seebeck measurement shows the absence of hopping conduction but potential fluctuations are present. Hall effect indicates inter-band conduction as observed by Oishi et al. [27].

(ii) High-temperature range: Here Seebeck measurement indicates a secondary transport mechanism like hopping might be present. Also, carrier concentration increases to degenerate level for $\beta$-Ga$_2$O$_3$.

Above two distinct conduction processes, can be explained if we assume a dominant defect interband is present in bandgap at low temperature which relatively suppressed at higher temperatures. Such defect bands in the bandgap of $\beta$-Ga$_2$O$_3$ have been confirmed by electrical and optical characterizations [27]. The spectral response characterization shows defect bands at 2.57 eV, 2.36 eV and 1.7 eV [2]. This solidifies our assumption that interband conduction is present in $\beta$-Ga$_2$O$_3$ at low temperatures. Moreover, a relatively low magnitude of the Seebeck coefficient (68.57 μV/K at 300 K) observed could not be explained if pure band conduction mechanism was present at low temperatures (< 300 K).

To summarize, thermoelectrical measurements of single-crystal $\beta$-Ga$_2$O$_3$ have been carried out in a wide temperature range (80-630K). The Hall effect and spectral response measurements have been used for the interpretation of observed behavior. Large and negative Seebeck coefficients suggest the electrons are dominant charge carriers in contributions to the transport properties. Based on the temperature-dependent Hall-mobility data analysis, a strong polar scattering effect has been found to be more prominent around room temperature. The Raman characterization showed the presence of low, mid, and high-frequency modes. The activation energy estimated from both the Seebeck coefficient and Hall-concentration data revealed that potential fluctuation and interband conduction might be present for low temperatures. The calculated power factor was found as ~ 0.3 μW/K$^2$ cm at 350 K and the *figure of merit* is ~0.01 at room temperature. Power factor behavior shows an increasing tendency with temperature suggesting suitability of *β*-Ga$_2$O$_3$ for thermoelectric material at higher temperatures. More studies are needed for better understanding of mechanism involved and enhancement of figure of merit in this material.

**Acknowledgments:** AK would like to acknowledge the funding received from the Department of Science, India through the INSPIRE scheme.

**Data Availability Statement:** The data that support the findings of this study are available from the corresponding author upon reasonable request.